\documentclass[twocolumn,aps,prc,showpacs,superscriptaddress,preprintnumbers,floatfix,nofootinbib]{revtex4}
\usepackage{epsfig,graphics}
\usepackage{graphicx}
\usepackage{dcolumn}
\usepackage{bm}
\usepackage{amsmath}
\usepackage[usenames]{color}
\usepackage{ulem} 

\newcommand{ \srt  }{$\sqrt{s_{_{\rm NN}}}$}

\newcommand{ \pt } {${p_{T}}$}

\newcommand{ \rdau } {${R_{dAu}}$}
\newcommand{\rts}   {${\sqrt{s_{NN}}}$}

\usepackage{CJK}
\usepackage{epstopdf}
\hfuzz=\maxdimen
\begin{document}
\begin{CJK*}{GB}{gbsn}

\title{$\phi$-meson production at forward/backward rapidity in high-energy nuclear collisions from a multiphase transport model}
\author{Y. J. Ye}
\affiliation{Shanghai Institute of Applied Physics, Chinese Academy of Sciences, Shanghai 201800, China}
\affiliation{University of Chinese Academy of Sciences, Beijing 100049, China}
\author{J. H. Chen}
\email{chenjinhui@sinap.ac.cn}
\affiliation{Shanghai Institute of Applied Physics, Chinese Academy of Sciences, Shanghai 201800, China}
\author{Y. G. Ma}
\email{ygma@sinap.ac.cn}
\affiliation{Shanghai Institute of Applied Physics, Chinese Academy of Sciences, Shanghai 201800, China}
\affiliation{ShanghaiTech University, Shanghai 200031, China}
\author{S. Zhang}
\affiliation{Shanghai Institute of Applied Physics, Chinese Academy of Sciences, Shanghai 201800, China}
\author{C. Zhong}
\affiliation{Shanghai Institute of Applied Physics, Chinese Academy of Sciences, Shanghai 201800, China}

\date{\today}

\begin{abstract}
Within the framework of a multiphase transport model (AMPT), the $\phi$-meson production is studied in d+Au collisions at \srt = {200} GeV in the forward  (d-going, $1.2<y<2.2$) and backward (Au-going, $-2.2<y<-1.2$) direction.
The AMPT model with string melting version (parton cascade turning-on) describes the experimental data well, while the pure hadronic transport scenario of the AMPT model underestimates the $\phi$-meson production rate in comparison with the data.
Detailed investigations including the rapidity, transverse momentum and collision system size dependencies of $\phi$-meson nuclear modification factor indicate that a combination of the initial state effect and a follow-up parton cascade is required in the AMPT model to describe the data. Similar calculations are also present in p+Pb collisions at \srt = {5.02} TeV and p+p collisions at \srt = {2.76} TeV. The findings from a comparison of AMPT model study with the data are consistent with that at RHIC energy.

 \end{abstract}

\maketitle

\section{INTRODUCTION}

Relativistic heavy-ion collisions provide a vital tool to mimic the matter of the early Universe at microseconds after the big bang. It is believed that this kind of matter is at a deconfined quark-gluon partonic state~\cite{RHIC-white-paper1,RHIC-white-paper2,RHIC-white-paper3,RHIC-white-paper4,Liu-NST,Ko-NST}. One of the important experimental methods is so-called nuclear modification factor which helps to understand the partonic matter created in central nucleus-nucleus collisions at the Relativistic Heavy Ion Collider (RHIC). In particularly, the observed strong suppression of identified particle spectra at high $p_{T}$  in Au+Au collisions with respect to the data in p+p collisions, i.e. so-called the jet quenching phenomenon, is attributed to parton energy loss when particle traverses through the hot medium~\cite{RHIC-white-paper1,RHIC-white-paper2,RHIC-white-paper3,RHIC-white-paper4,Jet-quenching,Nie-NT}. Intriguing phenomenon on stronger $J/\Psi$ suppression at forward rapidity in comparison to the result at middle rapidity has been observed and is interpreted as a combined contribution from hot matter effect and cold nuclear matter effect~\cite{PHENIX-JPsi}. Recently, experimental data from d+Au collisions at RHIC and from p+Pb collisions at LHC show that nuclear modification factor of $\phi$ mesons at forward rapidity is strongly suppressed in the d (p) going direction than the Au (Pb) going direction~\cite{PHENIX_phi_dau,ALICE-phi}. And the theoretical understanding of the new data at forward rapidity is limited. For example, how large fraction of the cold nuclear matter effect contributes to the data, or if the cold nuclear matter effect and the hot matter effect can be factorised, is not clear. In this paper, we use a multiphase transport (AMPT) model~\cite{AMPT-model} to investigate the dynamic process of $\phi$-meson production in d+Au collisions at the RHIC and p+Pb collisions at the LHC. We find that a parton cross section of 1.5$mb$ applied in the AMPT model describes the $\phi$-meson spectra at forward/backward rapidity well. The current study is different from the previous study on the identified particle production in d+Au collisions in the mid-rapidity region, where final state interaction plays a dominated role in the $p_T$ spectra of  $\pi$, K and p \cite{Xiaoping-dau}.

The paper is organized as follows. A brief description of the AMPT model is introduced in Sec. II. The results and discussion are presented  in Sec. III. Finally, a summary is given in Sec. IV.

\section{BRIEF DESCRIPTION OF THE AMPT MODEL}

The AMPT model is a hybrid model including the following four main components \cite{AMPT-model}: the initial condition, the partonic interactions, the conversion from partonic matter into hadronic matter and the hadronic interactions.
The initial condition, which includes the spatial and momentum distributions of minijet partons and soft string excitation, are obtained from the HIJING model \cite{HIJING}. Scattering among partons are modelled by Zhang's parton cascade (ZPC) \cite{ZPC}, which at present includes only two-body scattering with cross sections obtained from the pQCD with screening masses. In the default AMPT model, partons are recombined with their parent strings when they stop interaction, and the resulting strings are converted to hadrons using a Lund string fragmentation model \cite{Lund}. In the AMPT model with string melting, a simple quark coalescence model based on the quark spatial information is used to combine parton into hadrons.  The dynamics of the subsequent hadronic matter is then described by A Relativistic Transport (ART) model \cite{ART}. The details of the AMPT model can be found in Ref.~\cite{AMPT-model}. In the present study, we adopt the version of AMPT-v1.26-v2.26 with the default Lund string fragmentation parameters $\alpha=0.5$ and $b=0.9$~GeV$^{-2}$ in the HIJING model, the QCD coupling constant $\alpha_{s} = 0.33$, and the screening mass $\mu = 3.2$~fm$^{-1}$ to obtain a parton scattering cross section of 1.5 mb in the ZPC. The new parameters were tabulated in Ref.~\cite{Jun-AMPT} that are able to describe both the charged particle multiplicity density and the elliptic flow measured in heavy ion collisions at RHIC.

\section{RESULTS AND DISCUSSION}

\subsection{$\phi$ meson production in d+Au and p+p collisions at $\sqrt{s_{NN}}$=200 GeV}

\begin{figure}[htbp]
      \includegraphics[width=0.52\textwidth]{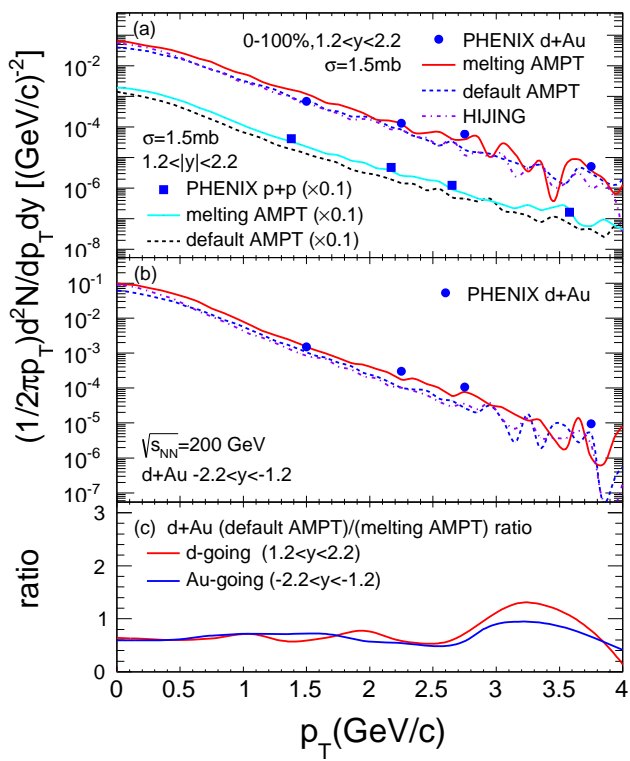}\\
      \caption{(Color online) (a)  Invariant yields of  $\phi$ mesons as a function of \pt~in the d-going direction in d+Au collisions and the results from p+p collisions at \rts = {200} GeV. (b) Invariant yields of  $\phi$ mesons as a function of \pt~ in the Au-going direction. Experimental data from PHENIX Collaboration are also plotted for comparison. (c) Ratio of the results from the AMPT model with string melting version divided by the results from the AMPT model with the default version.}
      \label{yield_dau}
    \end{figure}

The $\phi$ meson invariant yields from the AMPT model in  d+Au collisions at \rts = {200} GeV as a function of \pt~in the d-going direction ($1.2<y<2.2$) and in the Au-going region ($-2.2<y<-1.2$) are shown in Fig.\ref{yield_dau}. In the d-going side, the AMPT model with string melting scenario (version 2.26 for this study) describes the data well, while the default AMPT (version 1.26) underestimates the data by about 40\%. In the Au-going side, the AMPT model with string melting version describes the data well up till $p_T$=1.5 GeV/c and underestimates the yield in the higher $p_T$ region as shown in Fig.\ref{yield_dau} (b), which may be due to the small current quark masses used in the AMPT model so that partons are less affected by the radial flow effect~\cite{AMPT-model}. The default AMPT version underestimates the data in both d-going and Au-going directions. The current parameter set of AMPT model with string melting reproduces the p+p data perfectly while the default version underestimates the data. The ratio of $\phi$ meson invariant yield versus $p_T$ between the AMPT model with string melting version and default AMPT model in d+Au collisions at \rts = {200} GeV is shown in  Fig.\ref{yield_dau}(c). The ratio is close to 0.6 and is independent of $p_T$ within statistical uncertainty. In order to understand the origin of the difference between two scenarios of the AMPT model, we calculate the $\phi$ meson yield from HIJING (version 1.383 for this study). It is seen from Fig.~\ref{yield_dau} that the $p_T$ spectrum of $\phi$ meson from HIJING is softer than the experimental data. The results from HIJING are close to the results from the default AMPT model as shown in Fig.~\ref{yield_dau}(a) and (b), which may be due to the reason that the hadronic cross section of $\phi$ meson is small in high energy nuclear collisions. It seems that the difference of $\phi$ meson $p_T$ spectra between the string melting version and the default version is from the partonic interaction.

\subsection{Nuclear modification factor}

 \begin{figure}[htbp]
      \includegraphics[width=0.52\textwidth]{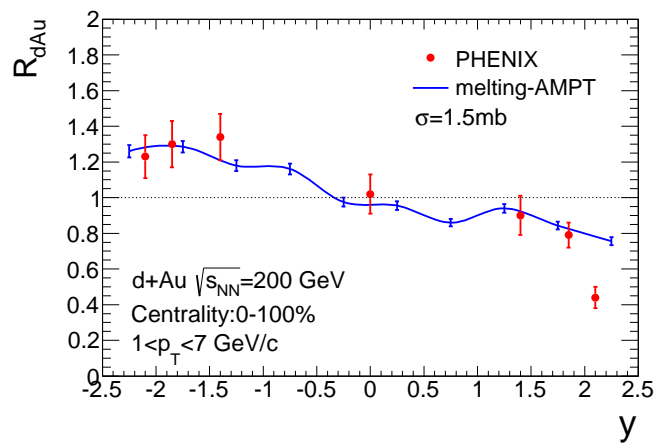}\\
      \caption{(Color online) The nuclear modification factor of $\phi$ meson as a function of rapidity. The line presents the results from the AMPT model with string melting scenario, while the solid circle data points show the experimental data from the PHENIX Collaboration~\cite{PHENIX_phi_dau}. }
      \label{Raa_y_melting_compare}
    \end{figure}

In order to explore  particle production mechanism and the nuclear medium effect in d+Au collisions, the ratio of the $\phi$ meson between the yield in d+Au collisions to p+p collisions scaled by the number of nucleon-nucleon collisions in the d+Au system, $N_{coll}$, is calculated as:
    \begin{equation}
        R_{dAu}=\frac{d^{2}N_{dAu}/dydp_{T}}{N_{coll}\times d^{2}N_{pp}/dydp_{T}},
        \label{eq10}
    \end{equation}
where $d^{2}N_{dAu}/dydp_{T}$ is the per-event yield of particle production in d+Au collisions and $d^{2}N_{pp}/dydp_{T}$ is the per-event yield of the same process in the p+p collisions. Figure~\ref{Raa_y_melting_compare} shows the \rdau~as a function of rapidity, summed over the \pt~range ($1<$\pt$<7$ GeV/c) and integrated over all centralities from the AMPT model with string melting scenario. The nuclear modification of $\phi$ production is enhanced in the Au-going direction and suppressed in the d-going direction from the AMPT model, and at middle rapidity ($|y| < 0.35$), the \rdau~is consistent with unit 1. The experimental results are also plotted in the figure. From the comparison, the AMPT model with string melting describes the data well in the measured rapidity range.

 \begin{figure}[htbp]
      \includegraphics[width=0.52\textwidth]{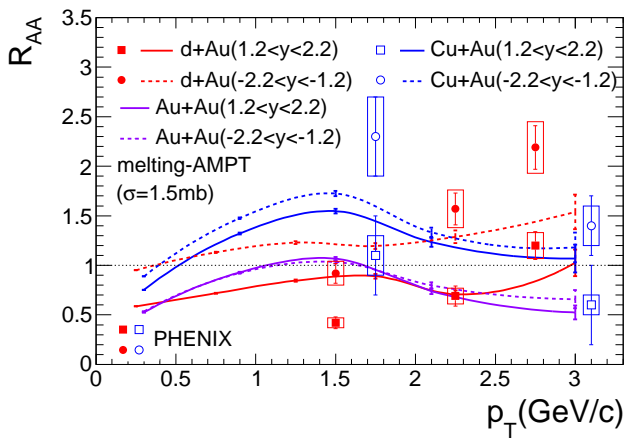}\\
      \caption{(Color online) The $R_{AA}$~as a function of \pt~from d+Au, Cu+Au and Au+Au collisions at \rts = {200} GeV. Lines represent results from the AMPT model with string melting version and data points are experimental results from the PHENIX Collaboration.}
      \label{Raa_CNM}
    \end{figure}

Cold nuclear matter (CNM) effect and hot matter effect are present together in heavy-ion collisions and both are important. The CNM effects usually incorporate the nuclear shadowing~\cite{Shadowing}, the Cronin enhancement~\cite{Cronin-effect}, and the initial state parton energy loss~\cite{Initial-state-E-loss}. The Cronin effect refers to the enhancement of high \pt~particle production in p+A collisions relative to that in p+p collisions scaled by the number of binary collisions and is attributed to partons undergoing multiple scattering within the nucleus~\cite{Cronin-effect}. Figure~\ref{Raa_CNM} shows the $\phi$ meson nuclear modification factor as a function of \pt~in the d-going and Au-going directions in d+Au collisions in comparison with the results from Cu+Au and Au+Au collisions at \rts = {200} GeV. In d+Au collisions, the \rdau~ increases with the increasing of transverse momentum and the yield of $\phi$ meson is enhanced at high \pt~with respect to the results from p+p collisions, which depicts the CNM effect in $\phi$ meson production in the forward/backward rapidity region at RHIC. For the Au-going direction, the \rdau~ shows an overall enhancement in comparison with d-going side, which stems from stronger multiple partonic scattering due to the larger size  of Au nucleus.  In the Au+Au and Cu+Au collisions, the $\phi$ meson $R_{AA}$ increases at low $p_T$ and starts to decrease at $p_T > $1.2 GeV/c as shown in the Fig.~\ref{Raa_CNM}. The difference on $p_{T}$ dependence of $\phi$ meson $R_{AA}$ between d+Au collisions and Au+Au (or Cu+Au) collisions may arise from the stronger hot nuclear matter effect in the dense medium created in Au+Au collisions at RHIC. In comparison with the experimental data from the PHENIX Collaboration in d+Au and Cu+Au collisions at~\rts = 200 GeV~\cite{PHENIX_phi_dau,PHENIX_phi_cuau}, the AMPT model with string melting scenario describes the~\pt~dependence of $\phi$ meson nuclear modification at forward rapidity reasonably well.

\subsection{Final state interaction effect on nuclear modification factor}

    \begin{figure}[htbp]
      \includegraphics[width=0.52\textwidth]{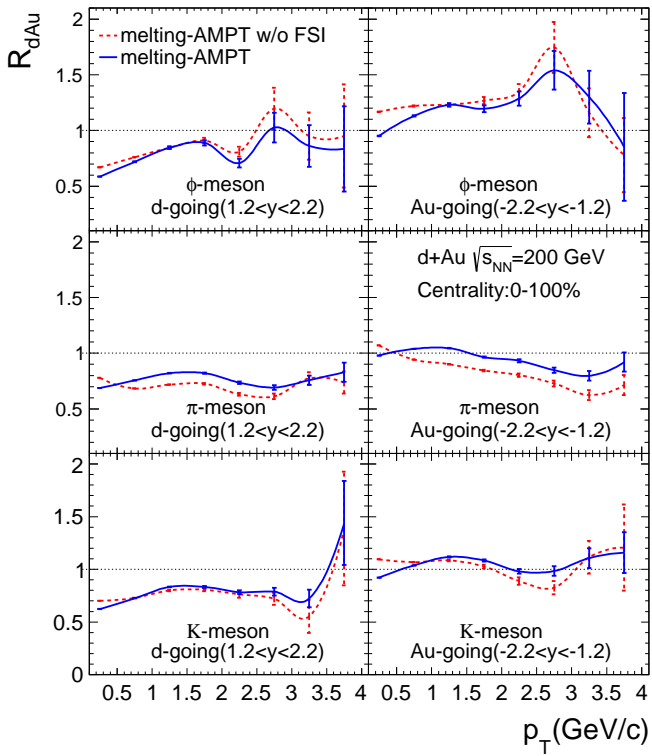}\\
      \caption{(Color online) The \rdau~of $\phi$ (top), $\pi$ (middle) and K (bottom) mesons as a function of \pt~in the forward rapidity region (left) and backward rapidity region (right) from the AMPT model with string melting scenario (solid lines). Results from the AMPT model without final state interactions are also plotted to address the FSI contributions (dash lines).}
      \label{Raa_pt_art}
    \end{figure}

Figure~\ref{Raa_pt_art} shows the \rdau~of $\phi$, $\pi$ and K meson as a function of \pt~at forward (left panels)  and backward (right panels) rapidities in d+Au collisions at \rts = {200} GeV from the AMPT model. The results from AMPT with string melting version without final state interaction (turned off the ART part) are also plotted for comparison. Top panels of Fig.~\ref{Raa_pt_art} show that the $\phi$ meson $R_{dAu}$ versus $p_T$ from the AMPT with and without final state interaction is close to each other and enhanced as pt increased at $p_{T}<3.0$ GeV/c. It may be due to the fact that $\phi$ meson hadronic cross section is small in high energy nuclear collisions. The enhancement of \rdau~versus $p_T$ could be from the Cronin effect as well. Middle panels of Fig.~\ref{Raa_pt_art} show that the value of the \rdau~of $\pi$ meson is larger in the AMPT model with FSI process turning-on, which may be due to the strong final state interaction and resonance decays in d+Au collisions in comparison to the p+p collisions. Bottom panels of Fig.~\ref{Raa_pt_art} depict the \rdau~of K meson. The results regarding to the FSI on Kaon \rdau~are similar to the results on the $\phi$ meson's. We learned that Lin and Ko have done a study~\cite{Lin-dAu} on the global properties of identified particles production in d+Au collisions at RHIC energy using the default AMPT model. The effect of final state interactions on the charged particle $p_T$ spectra in d+Au collisions is much smaller than observed in experimental data~\cite{Lin-dAu}.

From this phenomenological analysis, it is suggested that the  mean free path of $\phi$ mesons in hadronic medium is large because of its small cross section of scattering with hadrons, which is similar with the K meson~\cite{STAR-phi}. In this case, the final state interaction has no significant influence on the production rate of $\phi$ and K meson in d+Au collisions at \rts = {200} GeV. However, for the $\pi$ mesons, the yields are enhanced at both d-going and Au-going directions because of the large hadronic cross section and resonance strong decays.

\subsection{$\phi$ meson production at LHC energy}
\begin{figure}[htbp]
      \includegraphics[width=0.52\textwidth]{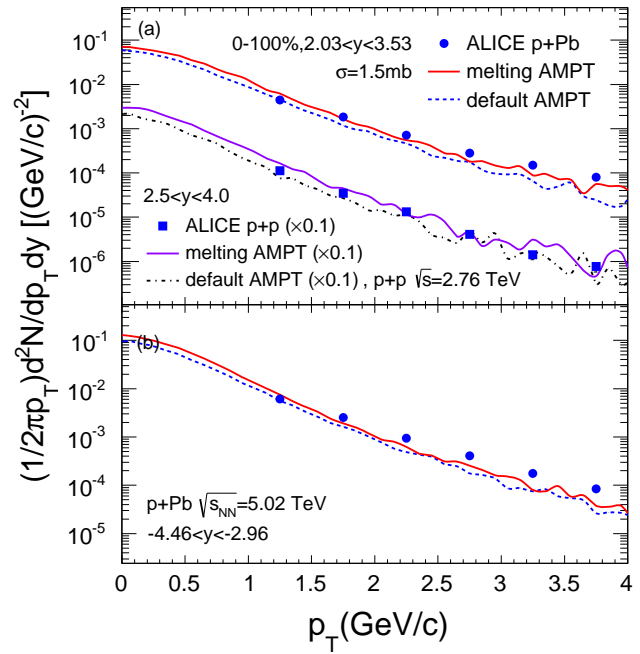}\\
      \caption{(Color online) (a)  Invariant yields of  $\phi$ mesons as a function of \pt~in the p-going direction in p+Pb collisions at \rts = {5.02} TeV and the results from p+p collisions at \rts = {2.76} TeV. (b) Invariant yields of  $\phi$ mesons as a function of \pt~in the Pb-going direction. Curves represent the results from the AMPT model, while data points are experimental results from the ALICE Collaboration~\cite{ALICE-phi}.}
      \label{yield_pPb}
    \end{figure}

The $\phi$ meson $p_T$ spectra in p+Pb collisions at \rts = {5.02} TeV and in p+p collisions at \rts = {2.76} TeV from the AMPT model are presented in Fig.~\ref{yield_pPb}. In the p-going direction ($2.03<y<3.53$), the AMPT model with string melting version describes the data reasonably well while the default AMPT model underestimates the data. In the Pb-going direction ($-4.46<y<-2.96$), the AMPT model with string melting describes the data up to $p_T$ = 1.7 GeV/c while systematically underpredicts the data at high $p_T$, which could be due to the small current quark masses used in the model~\cite{AMPT-model}. In p+p collisions, the AMPT model with string melting version describes the data in the whole $p_T$ region, while the default version underestimates the $\phi$ meson data. The feature is similar as observed in d+Au collisions at RHIC energy as described in Sec. III. A.

\section{Summary}

The $\phi$ meson productions at forward/backward rapidities in high energy nuclear collisions have been studied in a framework of multiphase transport model (AMPT). At RHIC energy, the AMPT model with string melting scenario describes the experimental data reasonably well for d+Au collisions at \rts = {200} GeV in the d-going direction, while it underestimates the data in higher $p_T$ region for the Au-going direction. The  default AMPT model underestimates the production rate by about 40\% in the covered $p_T$ range in comparison with the version with string melting scenario. The rapidity dependence of $\phi$ meson $R_{dAu}$  is well reproduced in the current AMPT calculation with  string melting scenario, which suggests that sufficient partonic interaction is required in order to describe the d + Au data. In the Au or Cu going direction, multiple partonic scattering is violent in contrast with the deuteron going direction, which results in an enhanced  $R_{dAu}$ in the Au or Cu going direction due to the size effect. The system size dependence of $\phi$ meson $R_{AA}$ versus transverse momentum shows the evolution from cold nuclear matter effect to hot nuclear matter effect when the size of system becomes larger.  In addition, our study on final-state hadronic interaction shows small contribution on the $\phi$ meson $R_{dAu}$. The $\phi$ meson $p_T$ spectra in p+Pb and p+p collisions at the LHC energy is also studied. The physics findings from the comparison of data with the AMPT model study are consistent with that at the RHIC energy.

\vspace{.5cm}

Valuable discussion with Prof. X. C. He from George State University are grateful. This work was supported in part by the Major State Basic Research Development Program in China under Contract Nos. 2014CB845400 and 2015CB856904, the National Natural Science Foundation of China under contract Nos. 11421505, 11520101004, 11322547 and 11275250.

\end{CJK*}
\end{document}